# Reconfigurable soliton networks optically-induced by arrays of nondiffracting Bessel beams

Zhiyong Xu, Yaroslav V. Kartashov, and Lluis Torner

*ICFO-Institut de Ciencies Fotoniques, and Department of Signal Theory and Communications, Universitat Politecnica de Catalunya, 08034 Barcelona, Spain*

*Yaroslav.Kartashov@icfo.es*

**Abstract:** We address the propagation of solitons in reconfigurable two-dimensional networks induced optically by arrays of nondiffracting Bessel beams in Kerr-type nonlinear media. We show that broad soliton beams can move across networks having different topologies almost without radiation losses, opening new prospects for all-optical soliton manipulation. We also discuss various switching scenarios for solitons launched into multi-core directional couplers optically-induced by suitable arrays of Bessel beams.

## 1. Introduction

Since their early theoretical prediction [1] and recent experimental observation [2,3], discrete optical solitons in waveguide arrays have attracted large attention because of their potential for all-optical switching and power- and angle-controlled steering (for a recent review see Ref. [4]). In particular, the possibility of construction of two-dimensional soliton networks of nonlinear waveguide arrays was established [5,6]. In landmark recent experiments [7-12] it was shown that periodic nonlinear lattices with flexibly controlled refractive index modulation depth and period can be induced all-optically, in particular, in photorefractive media. Such lattices constituted by continuous nonlinear media with an optically imprinted modulation of refractive index offer a number of new opportunities for the all-optical manipulation of light as well [13-16], since they can operate in both weak- and strong-coupling regimes, depending on the depth of refractive index modulation.

The technique of optical lattice induction opens a wealth of opportunities for creation of waveguiding configurations with various nondiffracting light beams. An important example is set by Bessel beams, which under ideal conditions do not diffract upon propagation. Single-mode Bessel beams allow creation of optically-induced lattices with radial symmetry, where solitons can be set into controllable rotary motion [17,18]. Combination of several incoherent Bessel beams can be used to build couplers and switching junctions that can trap and steer solitons, while junction properties can be tuned by intensity, intersection angles, and width of the central cores of Bessel beams. Notice that complex configurations of Bessel beams can be created, e.g., by computer-generated holograms [19-21].

In this paper we address different types of two-dimensional networks created with arrays of mutually incoherent parallel Bessel beams in Kerr-type nonlinear media. Each of the Bessel beams forming the network induces a well-pronounced guiding channel that overlaps with its neighbors through slowly decaying tails. We show that broad solitons launched across the network can jump between neighboring channels almost without radiation losses and thus can follow network bends, phenomena that open a wealth of opportunities for managing the soliton propagation trajectories. The networks suggested here are advantageous in comparison with their technologically prefabricated counterparts, because of the tunability afforded by the optical induction.

## 2. Model

We consider the propagation of light along the $\xi$ axis of a focusing Kerr-type nonlinear media with an optically-induced modulation of the refractive index in the transverse direction. The evolution of the complex light field amplitude $q$ is described by the reduced equation

$$i\frac{\partial q}{\partial \xi} = -\frac{1}{2}\left(\frac{\partial^2 q}{\partial \eta^2} + \frac{\partial^2 q}{\partial \zeta^2}\right) - q|q|^2 - pR(\eta,\zeta)q, \qquad (1)$$

where the transverse $\eta,\zeta$ and the longitudinal $\xi$ coordinates are scaled to the characteristic beam width and diffraction length, respectively. We suppose here that the refractive index modulation is induced optically by the multiple incoherent zero-order Bessel beams, so that the refractive index profile features the total intensity of the interference pattern, as it occurs in photorefractive crystals. The guiding parameter $p$ is proportional to the refractive index modulation depth, which is assumed to be comparable to the nonlinear contribution of the refractive index. The function $R(\eta,\zeta)$ describing the refractive index modulation profile is given by

$$R(\eta,\zeta) = \sum_{k=1}^{N} J_0^2\{(2b_{\text{lin}})^{1/2}[(\eta - \eta_k)^2 + (\zeta - \zeta_k)^2]^{1/2}\}, \qquad (2)$$

where $N$ is the total number of beams in the array, the scaling parameter $b_{\text{lin}}$ defines the radii of rings of Bessel beams and here is taken small enough to ensure that the width of the central core of Bessel beams largely exceeds the wavelength; finally, $\eta_k$ and $\zeta_k$ are the coordinates of the Bessel beam centers. Note, that Eq. (1) admits several conserved quantities including the power or energy flow

$$U = \int_{-\infty}^{\infty}\int_{-\infty}^{\infty} |q|^2 \, d\eta d\zeta. \qquad (3)$$

The stationary solutions of Eq. (1) that propagate along the guiding channels of the network have the form $q(\xi,\eta,\zeta) = w(\eta,\zeta)\exp(ib\xi)$, where $w(\eta,\zeta)$ is a real function and $b$ is the real propagation constant. General families of soliton solutions are defined by the propagation constant $b$, and by guiding and scaling parameters $p$ and

$b_{\text{lin}}$. Since scaling transformation $q(\eta,\zeta,\xi,p) \to \chi q(\chi\eta,\chi\zeta,\chi^2\xi,\chi^2 p)$ can be applied to obtain various soliton families from a given one (notice that this also implies variation of the lattice), below we set $b_{\text{lin}} = 10$ and vary $b$ and $p$. Families of stationary solutions were obtained by solving Eq. (1) with a relaxation algorithm.

Although here we focus on the optical context, we stress that the concept we put forward here is expected to hold for Bose-Einstein condensates trapped in Bessel optical lattices.

## 3. Discussion

First, we address the simplest uniform *line* network created with an array of Bessel beams equally spaced along $\eta$-axis (Fig. 1(a)). Because of the mutual incoherence of Bessel beams in the network, the interference pattern between pronounced guiding channels is suppressed. Such network can support stationary solitons whose profiles are elongated along the $\eta$ direction. This is especially evident for low-power solitons that extend over several network channels (Fig. 1(b)). With increase of the propagation constant (or power) the soliton width decreases (Fig. 1(c)), so that in the limit $b \to \infty$ soliton occupies only one channel and its power approaches the critical value of the unstable soliton supported by uniform cubic media. Nevertheless, for the considered set of parameters the soliton power was found to be a monotonically growing function of propagation constant that implies stability of stationary soliton solutions in the entire domain of their existence according to the Vakhitov-Kolokolov criterion. Below we consider broad low-power solitons that can be set in motion across the network by imposing the initial linear phase tilt (or angle) $\exp(i\alpha\eta)$ onto the input field distribution. Such tilted beams are no more stationary solitons, but they can travel across the uniform network along the $\eta$ direction almost without the radiation losses (Fig. 1(d)), that otherwise unavoidably appear upon crossing of the guiding channels of the structure and that then lead to a fast trapping of high-power narrow solitons.

One of the interesting results found in this work is shown in Fig. 1(d): Removal of one of the Bessel beams from the array can cause deflection of soliton in the direction opposite to the input tilt. This process is accompanied by a strong beam reshaping in the deflection point, but the resulting beam shape and the modulus of the propagation angle are typically very close to the input ones. The soliton deflection at the network defect is a simplest example of operation accessible with optically-induced networks whose structure (open and closed paths) can be easily changed by blocking or switching of the individual Bessel beams from the array. The potential of such structures for creation of all-optical photonic circuits is readily apparent.

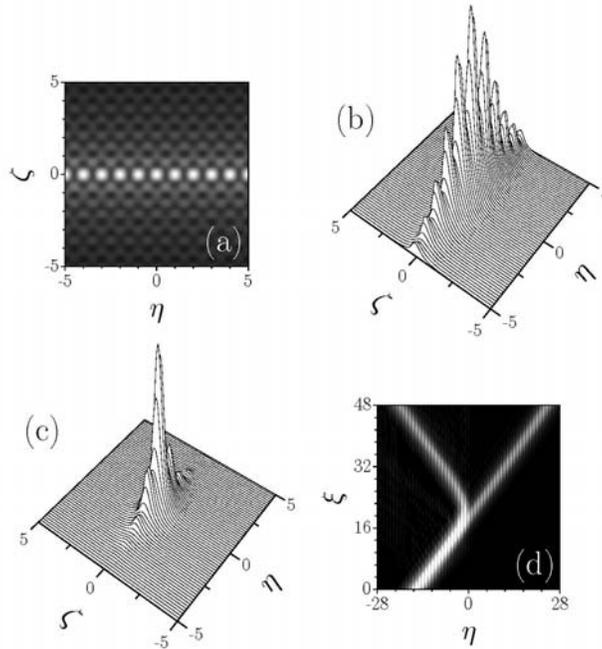

Fig. 1. (a) Linear network created with array of Bessel beams. Profiles of solitons with power $U = 0.8$ (b) and $U = 1.34$ (c) supported by network shown in (a). (d) Drift of soliton with power $U = 0.8$ and incident angle $\alpha = 1$ in uniform network and its deflection on the network defect. Intensity distributions showing drift and deflection are superimposed and taken at $\zeta = 0$. In all cases modulation depth $p = 15$ and separation between Bessel waveguides $\eta_0 = 1$.

Arrays of Bessel beams can be used to induce more complex networks that feature one or several bends at different angles that can even exceed 90°. Two representative examples of broad soliton propagation in such bent networks are shown in Figs. 2(a) and 2(b). Upon passing through the network, the soliton beams experience only slight transformation of their shape despite of the fact that they are forced to change propagation plane. The radiative losses in both cases shown in Fig. 2 are small (less than 4%). The networks shown in Figs. 2(a) and 2(b) are similar to the technologically fabricated networks studied in Refs [5,6]; however, the tunability afforded by the technique of optical-induction offer the additional advantage of the reconfigurability. Another interesting phenomenon occurs in the circular network, as the one depicted in Fig. 2(c). In this case soliton can perform rotary motion across the ring, with its power remaining almost constant because of small radiation losses. Switching off elements of the circular networks causes reversal of the soliton rotation direction. The similar results were obtained in a variety of cases, in terms of input light conditions and lattice-creating Bessel beams. These examples illustrate that the networks induced by the Bessel beam behave like *soliton wires* and thus they can be effectively used to manage soliton paths.

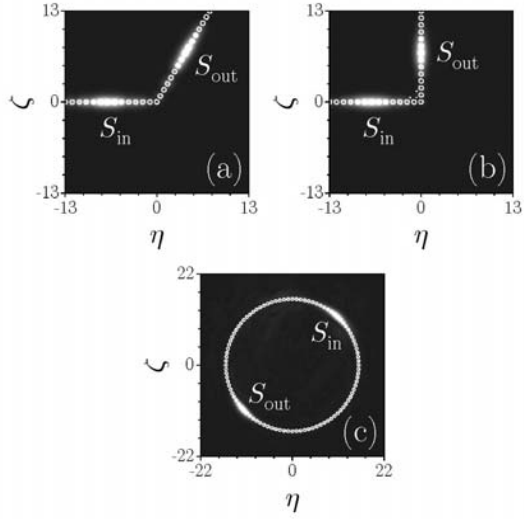

Fig. 2. Soliton propagation across 60°-bend (a), 90°-bend (b), and circular (c) networks created with arrays of Bessel beams. White contour lines are to help the eye and show positions of the induced guiding channels. Labels $S_{\text{in}}$, $S_{\text{out}}$ stand for input and output soliton positions. Input power $U = 0.8$, incident angle $\alpha = 1$, modulation depth $p = 15$, and separation between Bessel waveguides $\eta_0 = 1$.

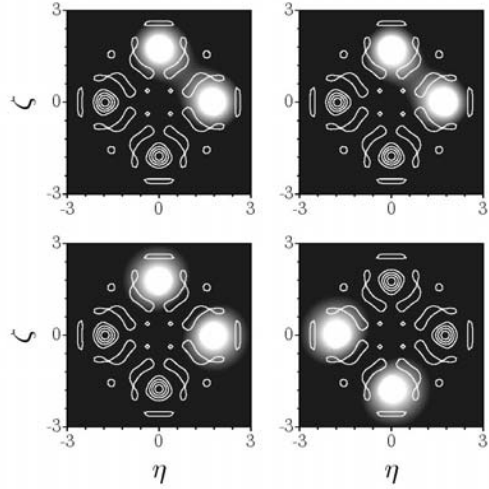

Fig. 3. Switching scenarios in the matrix of four Bessel waveguides when two solitons are launched into neighboring optically-induced guides. Top: input (left) and output (right) intensity distributions for in-phase solitons. Bottom: input (left) and output (right) intensity distributions for out-of-phase solitons. Input power $U = 2.14$, modulation depth $p = 5$, and separation between Bessel beams $\eta_0 = 2.5$.

Second, we study soliton switching in multi-core directional couplers created by arrays of incoherent Bessel beams. Two representative examples of the existing possibilities with four Bessel-beam couplers are shown in Figs. 3 and 4.

Here we focus on the results obtained by simultaneous excitation of several coupler channels. We launch into neighboring or opposite channels two identical solitons whose profiles were found from Eq. (1) by means of relaxation method, under the assumption that only one Bessel beam is present. In the presence of other guiding channels (with the second beam launched into one of them) the propagation process is accompanied by periodic transfer of the energy between different channels that are coupled via their tails. The energy exchange process is very sensitive to the relative phase of input beams. We show the output intensity distributions at the distance corresponding to the maximal energy transfer into two channels that were not initially occupied. Thus, in Fig. 3, where solitons are launched into neighboring channels, one finds that the energy transfer is negligible for in-phase solitons and almost 100% efficient for out-of-phase solitons. The outcome changes drastically when input solitons are launched into the opposite coupler channels. Therefore, the important result is that by varying the refractive index modulation depth (which is proportional to the intensity of nondiffracting Bessel beams inducing the coupler), the separation between Bessel channels, the power and phases of input solitons, it is possible to achieve a variety of switching scenarios.

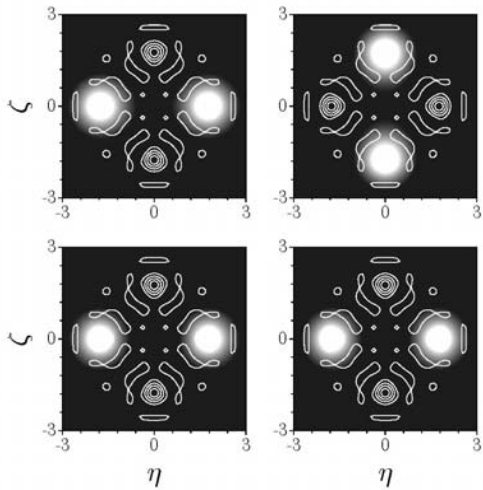

Fig. 4. Switching scenarios in the multicore coupler optically-induced by four Bessel beams when two solitons are launched into opposite waveguides. Top: input (left) and output (right) intensity distributions for in-phase solitons. Bottom: input (left) and output (right) intensity distributions for out-of-phase solitons. Input power $U = 2.14$, modulation depth $p = 5$, and separation between Bessel beams $\eta_0 = 2.5$.

## 4. Concluding remarks

We conclude stressing the potential of reconfigurable soliton networks optically induced by multiple incoherent nondiffracting Bessel beams in Kerr-type nonlinear media. We showed that dynamics of soliton beams propagating in such optically-induced networks can be used for all-optical manipulation of light [5,6], with the important additional advantage of the easy reconfigurability afforded the optical-induction concept. The scheme holds for light signals propagating in focusing cubic nonlinear media and for Bose-Einstein condensates trapped in optical lattices induced by Bessel beams.

**Acknowledgements**

This work has been partially supported by the Generalitat de Catalunya and by the Spanish Government through grant BFM2002-2861 and through the Ramon-y-Cajal program.

**References and links**